\documentclass[showpacs, pre]{revtex4}   

\usepackage{amsmath}
\usepackage{amssymb}
\usepackage{graphicx}
\usepackage{times}
\usepackage{verbatim}
\usepackage{algorithmic,algorithm}

\DeclareGraphicsExtensions{.eps}

\renewcommand{\baselinestretch}{2}       

\begin{document}

\title{Complex Networks Analysis of Combinatorial Spaces: the NK Landscape Case}
\author{Marco Tomassini}
\email{marco.tomassini@unil.ch}
\affiliation{Information Systems Institute, HEC, University of Lausanne, Switzerland}

\author{S\'ebastien V\'erel}
\email{verel@i3s.unice.fr}
\affiliation{University of Nice Sophia-Antipolis / CNRS, Nice, France\\ Institut des Syst\`emes Complexes, Paris Ile-de-France}

\author{Gabriela Ochoa}
\email{gxo@cs.nott.ac.uk} \affiliation{School of Computer Science,
University of Nottingham, Nottingham, UK}

\begin{abstract}

\noindent We propose a network characterization of combinatorial
fitness landscapes by adapting the notion of inherent networks
proposed for energy surfaces. We use the well-known family of $NK$
landscapes as an example. In our case the inherent network is the
graph whose vertices represent the local maxima in the landscape,
and the edges account for the transition probabilities between their
corresponding basins of attraction. We exhaustively extracted such
networks on representative $NK$ landscape instances, and performed a
statistical characterization of their properties. We found that most
of these network properties are related to the search difficulty on
the underlying $NK$ landscapes with varying values of $K$.

\end{abstract}


\maketitle

\section{Introduction}
\label{intro}

Difficult combinatorial landscapes are found in many important problems
in physics, computing, and in common everyday life activities such as resource allocation
and scheduling. For example, spin-glass systems give rise to
such energy landscapes which are characterized by many local
minima and high energy barriers between them. These landscapes generally
show frustration, i.e.~frozen disorder where the system is unable
to relax into a state in which all constraints are satisfied. In completely
different fields, such as combinatorial optimization, similar hard problems
also arise, for example the well-known \textit{traveling salesman problem}
and many others.

In order to understand the reasons that make these problems
difficult to optimize, a number of model landscapes have been
proposed. One of the simplest yet representative example is
Kauffman's family of $NK$ landsdcapes~\cite{kauffman93}. The $NK$
family of landscapes is a problem-independent model for constructing
multimodal landscapes that can gradually be tuned from smooth to
rugged, where the term ``rugged'' is intuitively related to the
degree of variability in the objective function value in neighboring
positions in configuration space. The more rugged the landscape, the
higher the number of local optima, and the landscape becomes
correspondingly more difficult to search for the global optimum. The
idea of an $NK$ landscape is to have $N$ ``spins" or ``genes'', each
with two possible values, $0$ or $1$. The model is a real stochastic
function $\Phi$ defined on binary strings $s \in  \{0,1\}^N$ of
length $N$, $\Phi: s \rightarrow \mathbb{R_+}$. The value of $K$
determines how many other spin values in the string influence a
given spin  $s_i, \: i=1,\ldots,N$. The value of $\Phi$ is the
average of the contributions $\phi_i$ of all the spins:

$$
\Phi(s) = \frac{1} {N} \sum_{i=1}^N \phi_i(s_i, s_{{i}_1}, \ldots, s_{{i}_K})
$$

By increasing the value of $K$ from 0 to $N-1$, $NK$ landscapes can
be tuned from smooth to rugged. For $K=0$ all contributions can be
optimized independently which makes $\Phi$ a simple additive
function with a single maximum. At the other extreme when $K=N-1$
the landscape becomes completely random, the probability of any
given configuration of being the optimum is $1/(N+1)$, and the
expected number of local optima is $2^N/(N+1)$. Intermediate values
of $K$ interpolate between these two cases and have a variable
degree of ``epistasis'', i.e. of spin (or gene)
interaction~\cite{kauffman93}.

The $K$ variables that form the context of the fitness contribution
of gene $s_i$ can be chosen according to different models. The two
most widely studied models are the {\em random neighborhood} model,
where the $K$  variables are chosen randomly according to a uniform
distribution among the $N-1$ variables other than $s_i$, and the
{\em adjacent neighborhood} model, in which the $K$ variables are
those closest to $s_i$ in a total ordering $s_1, s_2, \ldots, s_N$
(using periodic boundaries). No significant differences between the
two models were found in terms of  global properties of the
respective families of landscapes, such as mean number of local
optima or autocorrelation length \cite{weinberger91,kauffman93}.
Similarly, our preliminary studies on the characteristics of the
$NK$ landscape optima networks did not show noticeable differences
between the two neighborhood models. Therefore, we conducted our
full study on the more general random model.

The $NK$ model is related to spin glasses, and more precisely to
$p$-spin models~\cite{derrida81,gross-mezard-84}, where $p$ plays a role similar to
$K$. In spin glasses the function analogous to $\Phi$
is the energy $H$ and the stable states are the minima of the energy
hypersurface.

In this study we seek to provide fundamental new insights into the
structural organization of the local optima in combinatorial
landscapes, particularly into the connectivity of their basins of
attraction. Combinatorial landscapes can be seen as a graph whose
vertices are the possible configurations. If two configurations can
be transformed into each other by a suitable operator move, then we
can trace an edge between them. The resulting graph, with an
indication of the fitness at each vertex, is a representation of the
given problem fitness landscape. A useful a simplification of the
graphs for the energy landscapes of atomic clusters was introduced
in \cite{doye02,doye05}. The idea consists in taking as vertices of
the graph not all the possible configurations, but only those that
correspond to energy minima. For atomic clusters these are
well-known, at least for relatively small assemblages. Two minima
are considered connected, and thus an edge is traced between them,
if the energy barrier separating them is sufficiently low. In this
case there is a transition state, meaning that the system can jump
from one minimum to the other by thermal fluctuations going through
a saddle point in the energy hyper-surface. The values of these
activation energies are mostly known experimentally or can be
determined by simulation. In this way, a network can be built which
is called the ``inherent structure'' or ``inherent network''
in~\cite{stillinger,doye02}.

We propose a network characterization of combinatorial fitness
landscapes by adapting the notion of {\em inherent networks}
described above. We use the  family of $NK$ landscapes as an
example. In our case the inherent network is the graph where the
vertices are all the local maxima and the edges account for
transition probabilities between maxima. We exhaustively extract
such networks on representative small $NK$ landscape instances, and
perform a statistical characterisation of their properties. Our
analysis is inspired, in particular, by the work of
\cite{doye02,doye05} on energy landscapes, and in general, by the
field of complex networks~\cite{newman03}. A related work can be
found in~\cite{scala-et-al-01}, where the case of lattice polymer
chains is studied. However, the notion of an edge there is very
different, being related to moves that bring a given conformation
into an allowed neighboring one. Similar ideas have been put forward
in physical chemistry to understand the thermodynamics and kinetics
of complex biomolecules through the network study of their
free-energy landscapes~\cite{gfeller-07}. It should also be noted
that our approach is different from the barrier-tree representations
of landscapes proposed by Stadler et al. (see, for
example,~\cite{stadler-02}).

The next section describes how combinatorial landscapes are mapped
onto networks, and includes the relevant definitions and algorithms
used in our study. The empirical analysis of our selected $NK$
landscape instances is presented in the following two sections; one
devoted to the study of basins, and the other to the network
statistical features. Finally, we present our conclusions and ideas
for future work.

\section{Landscapes as Networks}
\label{landscapes}

Many natural and artificial systems can be modeled as networks.
Typical examples are communication systems (the Internet, the web,
telephonic networks), transportation lines (railway, airline
routes), biological systems (gene and protein interactions), and
social interactions (scientific co-authorship, friendships). It has
been shown in recent years that many of these networks exhibit was
has been called a {\em small-world} topology \cite{newman03}, in
which nodes are highly clustered yet the path length between them is
small. Additionally, in several of these networks the distribution
of the number of neighbours (the degree distribution) is typically
right-skewed with a ``heavy tail", meaning that most of the nodes
have less-than-average degree whilst a small fractions of hubs have
a large number of connections. These topological features are very
relevant since they impact strongly on networks' properties such as
their robustness and searchability.

To model a physical energy landscape as a network, Doye~\cite{doye02}
needed to decide first on a definition both of a state of the system
and how two states were connected. The states and their connections
will then provide the nodes and edges of the network. For systems
with continuous degrees of freedom, this was achieved through
the `inherent structure' mapping. In this mapping each point in
configuration space is associated with the minimum (or `inherent
structure') reached by following a steepest-descent path from that
point. This mapping divides configurations into basins of attraction
surrounding each minimum on the energy landscape.

Our goal is to adapt this idea to the context of combinatorial
optimization. In our case, the vertexes of the graph can be
straightforwardly defined as the local maxima of the landscape.
These maxima are obtained exhaustively by running a best-improvement
local search algorithm (see Fig. \ref{algoHC}) from every
configuration of the search space. The definition of the edges,
however, is a much more delicate matter. In our initial
attempt~\cite{gecco-nk-08} we considered that two maxima $i$ and $j$
were connected (with an undirected and unweighed edge), if there
exists at least one pair of direct neighbors solutions $s_i$ and
$s_j$, one in each basin of attraction ($b_i$ and $b_j$). We found
empirically on small instances of $NK$ landscapes, that such
definition produced densely connected graphs, with very low ($\leq
2$) average path length between nodes for all $K$. Therefore, apart
from the already known increase in the number of optima with
increasing $K$, no other network property accounted for the increase
in search difficulty. Furthermore, a single pair of neighbours
between adjacent basins, may not realistically account for actual
basin transitions occurring when using common heuristic search
algorithms. These considerations, motivated us to search for
alternative definitions of the edges connecting local optima. In
particular, we decided to associate weights to the edges that
account for the transition probabilities between the nodes (local
optima). More details on the relevant algorithms and formal
definitions are given below.

\subsection{Definitions and Algorithms}
\label{sec:defs}

\textbf{Definition:} Fitness landscape~\cite{stadler-02}.\\
A landscape is a triplet $(S, V, f)$ where $S$ is a set of potential
solutions i.e. a search space, $V : S \longrightarrow 2^S$, a
neighborhood structure, is a function that assigns to every $s \in
S$ a set of neighbors $V(s)$, and $f : S \longrightarrow R$ is a
fitness function that can be pictured as the \textit{height} of the
corresponding solutions.

In our study, the search space is composed by binary strings of
length $N$, therefore its size is $2^N$. The neighborhood is defined
by the minimum possible move on a binary search space, that is, the
1-move or bit-flip operation. In consequence, for any given string
$s$ of length $N$, the neighborhood size is $|V(s)| = N$.

The $HillClimbing$ algorithm to determine the local optima and
therefore define the basins of attraction, is given below:

\renewcommand{\baselinestretch}{1}       
\begin{figure}[!ht]
\begin{center}
\begin{algorithmic}
\STATE Choose initial solution $s \in S$ \REPEAT
    \STATE choose $s^{'} \in V(s)$ such that $f(s^{'}) = max_{x \in V(s)}\ f(x)$
        \IF{$f(s) < f(s^{'})$}
            \STATE $s \leftarrow s^{'}$
    \ENDIF
\UNTIL{$s$ is a  Local optimum}
\end{algorithmic}
\caption{{\em HillClimbing} algorithm.}
\label{algoHC}
\end{center}
\end{figure}

\renewcommand{\baselinestretch}{2}       

\textbf{Definition:} Local optimum.\\
A local optimum is a solution $s^{*}$ such that $\forall  s \in
V(s^{*})$, $f(s) < f(s^{*})$.

\noindent The $HillClimbing$ algorithm defines a mapping from the
search space $S$ to the set of locally optimal solutions $S^*$.\\

\textbf{Definition:} Basin of attraction.\\
The basin of attraction of a local optimum $i \in S$ is the set $b_i = \{
s \in S ~|~ HillClimbing(s) = i \}$. The size of the basin of
attraction of a local optimum $i$ is the cardinality of $b_i$.

Notice that for non-neutral fitness landscapes, as are $NK$
landscapes, the basins of attraction as defined above, produce a
partition of the configuration space $S$. Therefore, $S = \cup_{i
\in S^{*}} b_i$ and $\forall i \in S$ $\forall j \not= i$, $b_i \cap
b_j = \emptyset$\\

 \textbf{Definition:} Edge weight.\\
\noindent For each pair of solutions $s$ and $s^{'}$, let us define $p(s
\rightarrow s^{'} )$ as the probability
to pass from $s$ to $s^{'}$ with the given neighborhood structure.
In the case of binary strings of size $N$, and the neighborhood
defined by the single bit-flip operation, there are $N$ neighbors
for each solution, therefore:

\noindent if $s^{'} \in V(s)$ , $p(s \rightarrow s^{'} ) = \frac{1}{N}$ and \\
if $s^{'} \not\in V(s)$ , $p(s \rightarrow s^{'} ) = 0$.

\noindent We can now define the probability to pass from a solution $s
\in S$ to a solution belonging to the basin $b_j$, as:
$$
p(s \rightarrow b_j ) = \sum_{s^{'} \in b_j} p(s \rightarrow s^{'} )
$$

\noindent Notice that $p(s \rightarrow b_j ) \leq 1$.

\noindent Thus, the total probability of going from basin $b_i$ to
basin $b_j$ is the average over all $s \in b_i$ of the transition
probabilities  to solutions $s^{'} \in b_j$ :

$$p(b_i \rightarrow b_j) = \frac{1}{\sharp b_i} \sum_{s \in b_i} p(s \rightarrow b_j )$$

\noindent $\sharp b_i$ is the size of the basin $b_i$.
We are now prepared to define our `inherent' network or network of
local optima.\\

\textbf{Definition:} Local optima network.\\
The local optima network $G=(S^*,E)$ is the graph where the nodes
are the local optima, and there is an edge $e_{ij} \in E$ with
weight $w_{ij} = p(b_i \rightarrow b_j)$ between two nodes $i$ and
$j$ if $p(b_i \rightarrow b_j) > 0$.

Notice that since each maximum has its associated basin, $G$ also
describes the interconnection of basins.

According to our definition of edge weights, $w_{ij} = p(b_i
\rightarrow b_j)$ may be different than $w_{ji} = p(b_j \rightarrow
b_i)$. Thus, two weights are needed in general, and we have an
oriented transition graph.

The following two definitions are relevant to the discussion of the
boundary of basins.

\textbf{Definition:} Boundary of a basin of attraction.\\
The \textit{boundary} $B(b)$ of a basin of attraction $b$ can be
defined as the set of configurations within a basin that have at
least one neighbor's solution in another basin $b^{'}$.

\textbf{Definition:} Interior of a basin of attraction.\\
Conversely, the \textit{interior} $I(b)$ of a basin is composed by
the configurations that have all their neighbors in the same basin.
Formally,
$$
B(b) = \lbrace s \in b \; | \: \exists \: b^{'} \not= b, \: \exists \: s^{'} \in
b^{'}, \: \exists \: e_{ss^{'}} \in E \rbrace,
$$
\vspace{-1cm}

$$I(b) = b-B(b)
$$

\section{Empirical Analysis of Basins}
\label{analysis}

In order to avoid sampling problems that could bias the results, we
used the largest values of $N$ that can still be analyzed
exhaustively with reasonable computational resources. We thus
extracted the local optima networks of landscape instances with $N =
{14, 16, 18}$, and $K = {2,4,6,..., N-2,N-1}$. For each pair of $N$
and $K$ values, 30 randomly generated instances were explored.
Therefore, the networks statistics reported below represent the
average behaviour of 30 independent instances.

Besides the maxima network, it is useful to describe the associated
basins of attraction as these play a key role in search algorithms.
Furthermore, some characteristics of the basins can be related to
the optima network features. The notion of the basin of attraction
of a local maximum has been presented before. We have exhaustively
computed the size and number of all the basins of attraction for
$N=16$ and $N=18$ and for all even $K$ values plus $K=N-1$. In this
section, we analyze the basins of attraction from several points of
view as  described below.

\subsection{Global optimum basin size versus $K$}

\begin{figure} [!ht]
\begin{center}
\includegraphics[width=8cm] {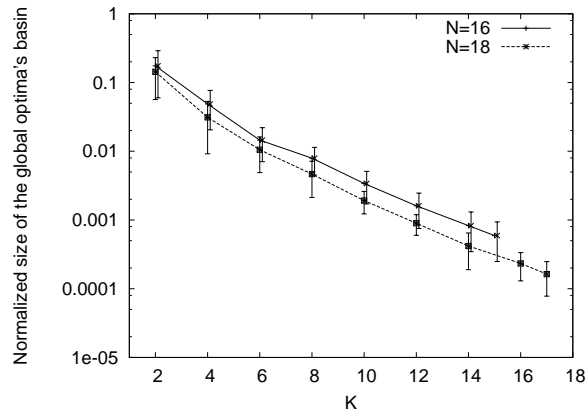} \protect \\
\caption{Average of the relative size of the basin corresponding to
the global maximum for each K over 30 landscapes.\label{nbasins}}
\end{center}
\end{figure}

In Fig.~\ref{nbasins} we plot the average size of the basin
corresponding to the global maximum for $N=16$ and $N=18$, and all
values of $K$ studied. The trend is clear: the basin shrinks very
quickly with increasing $K$. This confirms that the higher the $K$
value, the more difficult for an stochastic search algorithm to
locate the basin of attraction of the global optimum

\subsection{Number of basins of a given size}

\begin{figure} [!ht]
\begin{center}
    \mbox{\includegraphics[width=8cm]{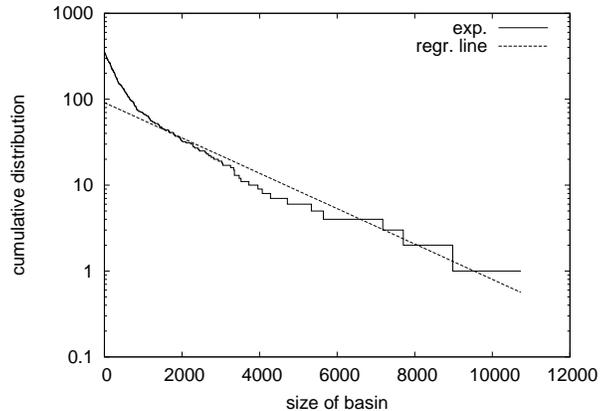} } \protect \\
\caption{Cumulative distribution of the number of basins of a given
size with regression line. A representative landscape with $N=18$, and $K=4$ is
visualized. A lin-log scale is used. \label{bas-18-size}}
\end{center}
\end{figure}

\begin{table}[!ht]
\begin{center}
\small \caption{Correlation coefficient ($\bar{\rho}$), and linear
regression coefficients (intercept ($\bar{\alpha})$ and slope
($\bar{\beta}$)) of the relationship between the basin size of
optima and the cumulative number of nodes of a given (basin) size (
in logarithmic scale: $\log(p(s)) = \alpha + \beta s + \epsilon$).
The average and standard deviation values over 30 instances, are
shown.}

\renewcommand{\baselinestretch}{1}       
\label{tab:cumulSize} \vspace{0.2cm}
\begin{tabular}{|c|c|c|c|}
\hline
\multicolumn{4}{|c|}{$N = 16$} \\
\hline
 $K$ & $\bar \rho$ & $\bar \alpha$ & $\bar{\beta}$ \\
\hline
2   & $-0.944_{0.0454}$ & $2.89_{0.673}$ & $-0.0003_{0.0002}$  \\
4   & $-0.959_{0.0310}$ & $4.19_{0.554}$ & $-0.0014_{0.0006}$  \\
6   & $-0.967_{0.0280}$ & $5.09_{0.504}$ & $-0.0036_{0.0010}$  \\
8   & $-0.982_{0.0116}$ & $5.97_{0.321}$ & $-0.0080_{0.0013}$  \\
10  & $-0.985_{0.0161}$ & $6.74_{0.392}$ & $-0.0163_{0.0025}$  \\
12  & $-0.990_{0.0088}$ & $7.47_{0.346}$ & $-0.0304_{0.0042}$  \\
14  & $-0.994_{0.0059}$ & $8.08_{0.241}$ & $-0.0508_{0.0048}$  \\
15  & $-0.995_{0.0044}$ & $8.37_{0.240}$ & $-0.0635_{0.0058}$  \\
\hline \hline
\multicolumn{4}{|c|}{$N = 18$} \\
\hline
2   & $-0.959_{0.0257}$ & $3.18_{0.696}$ & $-0.0001_{0.0001}$  \\
4   & $-0.960_{0.0409}$ & $4.57_{0.617}$ & $-0.0005_{0.0002}$  \\
6   & $-0.967_{0.0283}$ & $5.50_{0.520}$ & $-0.0015_{0.0004}$  \\
8   & $-0.977_{0.0238}$ & $6.44_{0.485}$ & $-0.0037_{0.0007}$  \\
10  & $-0.985_{0.0141}$ & $7.24_{0.372}$ & $-0.0077_{0.0011}$  \\
12  & $-0.989_{0.0129}$ & $7.98_{0.370}$ & $-0.0150_{0.0019}$  \\
14  & $-0.993_{0.0072}$ & $8.69_{0.276}$ & $-0.0272_{0.0024}$  \\
16  & $-0.995_{0.0056}$ & $9.33_{0.249}$ & $-0.0450_{0.0036}$  \\
17  & $-0.992_{0.0113}$ & $9.49_{0.386}$ & $-0.0544_{0.0058}$  \\
\hline
\end{tabular}
\end{center}
\end{table}
\renewcommand{\baselinestretch}{2}

Fig.~\ref{bas-18-size} shows the cumulative distribution of the
number of basins of a given size (with regression line) for a
representative instances with $N=18$, $K = 4$ . Table
~\ref{tab:cumulSize} shows the average (of 30 independent
landscapes) correlation coefficients and linear regression
coefficients (intercept ($\bar{\alpha})$ and slope ($\bar{\beta}$))
between the number of nodes and the basin sizes for
instances with $N=16, 18$ and all for all the studied values of $K$.
Notice that
distribution decays exponentially or faster for the lower $K$ and it
is closer to exponential for the higher $K$. This observation is
relevant to theoretical studies that estimate the size of attraction
basins (see for example \cite{garnier01}). These studies often
assume that the basin sizes are uniformly distributed, which is not
the case for the $NK$ landscapes studied here. From the slopes $\bar
\beta$ of the regression lines (table~\ref{tab:cumulSize}) one can
see that high values of $K$ give rise to steeper distributions
(higher $\bar \beta$ values). This indicates that there are less
basins of large size for large values of $K$. In consequence, basins
are broader for low values of $K$, which is consistent with the fact
that those landscapes are smoother.

\subsection{Fitness of local optima versus their basin sizes}

\begin{figure} [!ht]
\begin{center}
    \mbox{\includegraphics[width=8cm]{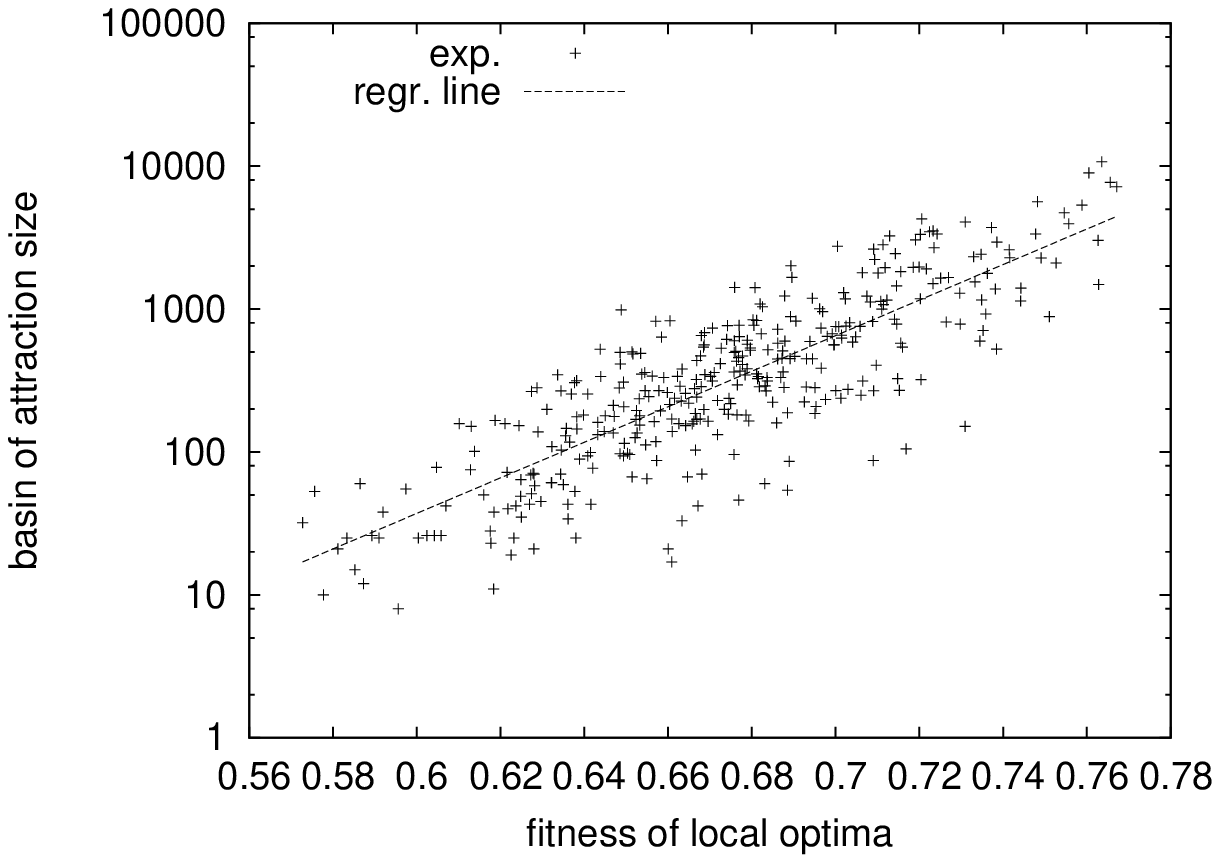} } \protect \\
    \mbox{\includegraphics[width=8cm]{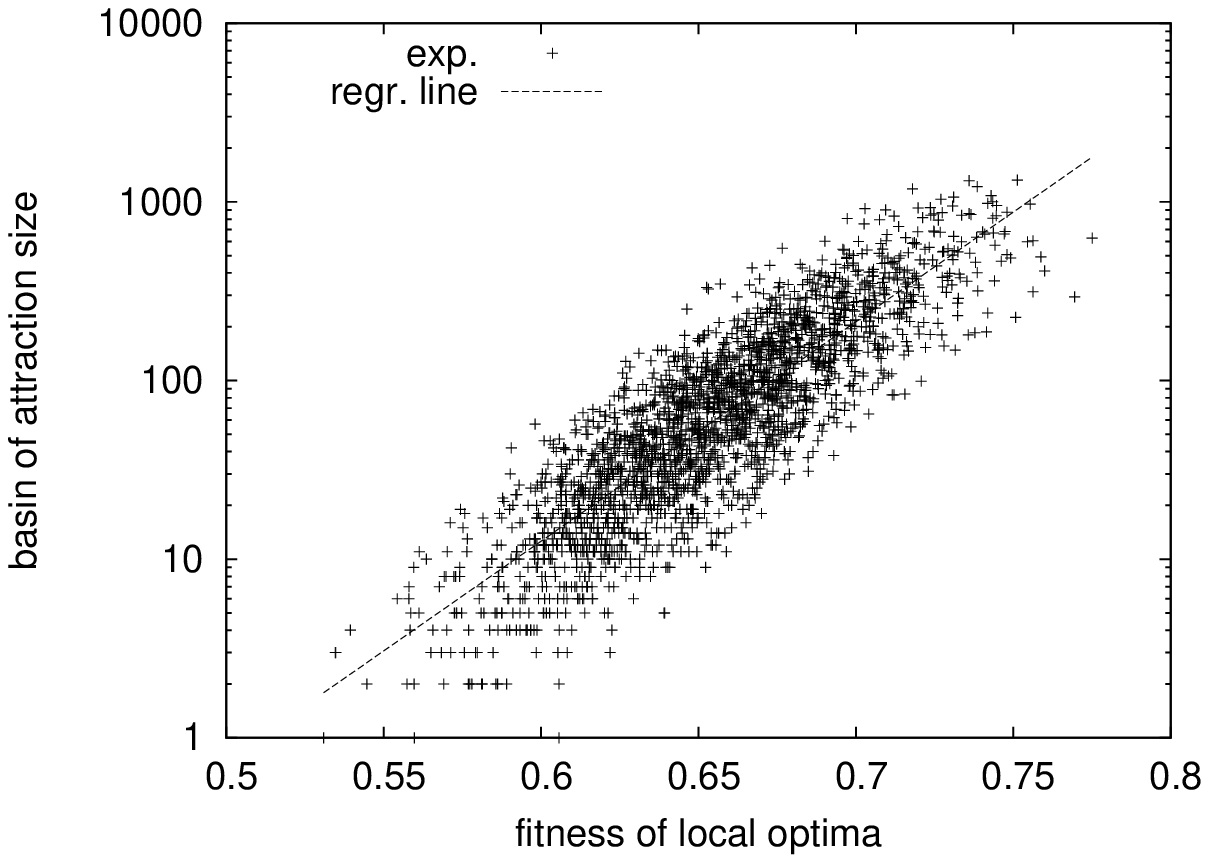} } \protect \\
\caption{Correlation between the fitness of local optima and their
corresponding basin sizes, for two representative instances with
$N=18$, $K=4$ (top) and $K=8$ (bottom). \label{fig:cor_fit-size}}
\end{center}
\end{figure}

The scatter-plots in Fig.~\ref{fig:cor_fit-size} illustrate the
correlation between the basin sizes of local maxima (in logarithmic
scale) and their fitness values. Two representative instances for
$N$ = 18 and $K$ = 4, 8 are shown. Notice that there is a clear
positive correlation between the fitness values of maxima and their
basins' sizes. In other words, the higher the peak the wider tend to
be its basin of attraction. Therefore, on average, with a stochastic
local search algorithm, the global optimum would be easier to find
than any other local optimum. This may seem surprising. But, we have
to keep in mind  that as the number of local optima increases (with
increasing $K$), the global optimum basin is more difficult to reach
by a stochastic local search algorithm (see Fig.~\ref{nbasins}).
This observation offers a mental picture of $NK$ landscapes: we can
consider the landscape as composed of a large number of mountains
(each corresponding to a basin of attraction), and those mountains
are wider the taller the hilltops. Moreover, the size of a mountain
basin grows exponentially with its hight.

\section{General Network Statistics}
\label{net-stat}

We now briefly describe the  statistical measures used for our
analysis of maxima networks.

The standard  clustering coefficient \citep{newman03} does not
consider weighted edges. We thus use the {\em weighted clustering}
measure proposed by \cite{bart05}, which combines the topological
information with the weight distribution of the network:

$$c^{w}(i) = \frac{1}{s_i(k_i - 1)} \sum_{j,h} \frac{w_{ij} + w_{ih}}{2} a_{ij} a_{jh} a_{hi}$$
where $s_i = \sum_{j \not= i} w_{ij}$, $a_{nm} = 1$ if $w_{nm} > 0$,
$a_{nm} = 0$ if $w_{nm} = 0$ and $k_i = \sum_{j \not= i} a_{ij}$.

For each triple formed in the neighborhood of the vertex $i$,
$c^{w}(i)$ counts the weight of the two participating edges of the
vertex $i$. $C^w$ is defined as the weighted clustering coefficient
averaged over all vertices of the network.

The standard topological characterization of networks is obtained by
the analysis of the probability distribution $p(k)$ that a randomly
chosen vertex has degree $k$. For our weighted networks, a
characterization of weights is obtained by the {\em connectivity and
weight distributions} $p_{in}(w)$ and $p_{out}(w)$ that any given edge has
incoming or outgoing weight $w$.

In our study, for each node $i$, the sum of outgoing edge weights
is equal to $1$ as they represent transition probabilities. So, an important
measure is the weight $w_{ii}$
of self-connecting edges (remaining in the same node). We have the
relation: $ w_{ii} + s_i = 1$. The vertex {\em strength}, $s_i$, is
defined as $s_i = \sum_{j \in {\cal V}(i) - \{i\}}  w_{ij}  $, where
the sum is over the set ${\cal V}(i) - \{i\}$ of neighbors of
$i$~\citep{bart05}. The strength of a node is a generalization of
the node's connectivity giving information about the number and
importance of the edges.

Another network measure we report here is {\em disparity}
\citep{bart05} $Y_{2}(i)$, which measures how heterogeneous are the
contributions of the edges of node $i$ to the total weight
(strength):

$$Y_{2}(i) = \sum_{j \not= i} \left( \frac{w_{ij}}{s_i} \right)^2 $$

The disparity could be averaged over the node with the same degree
$k$. If all weights are nearby of $s_i/k$, the disparity for nodes
of degree $k$ is nearby $1/k$.

Finally, in order to compute the average shortest path
between two nodes on the optima network of a given landscape, we
considered the expected number of bit-flip mutations to pass from
one basin to the other. This expected number can be computed by
considering the inverse of the transition probabilities between
basins. In other words, if we attach to the edges the inverse of the
transition probabilities, this value would represent the average
number of random mutations to pass from one basin to the other. More
formally, the distance (expected number of bit-flip mutations)
between two nodes is defined by $d_{ij} = 1 / w_{ij}$ where $w_{ij}
= p(b_i \rightarrow b_j)$. Now, we can define the length of a path
between two nodes as being the sum of these distances along the
edges that connect the respective basins.

\renewcommand{\baselinestretch}{1}
\begin{table}[!ht]
\begin{center}
\small \caption{$NK$ landscapes network properties.  Values are
averages over 30 random instances, standard deviations are shown as
subscripts. $n_v$ and $n_e$ represent the number of vertexes and
edges (rounded to the next integer), $\bar C^{w}$, the mean weighted
clustering coefficient. $\bar Y$ represent the mean disparity
coefficient, $\bar d$ the mean path length (see text for
definitions). } \label{tab:statistics} \vspace{0.2cm}

\begin{tabular}{|c|c|c|c|c|c|}
\hline
$K$ & $\bar n_v$ & $\bar n_e$ & $\bar C^{w}$ & $\bar Y$ & $\bar d$ \\
\hline
\multicolumn{6}{|c|}{$N = 14$} \\
\hline
2  &    $14_{ 6}$  & $200_{131}$      & $0.98_{0.0153}$ & $0.367_{0.0934}$ & $76_{194}$ \\
4  &    $70_{10}$  & $3163_{766}$     & $0.92_{0.0139}$ & $0.148_{0.0101}$ & $89_{6}$   \\
6  &   $184_{15}$  & $12327_{1238}$   & $0.79_{0.0149}$ & $0.093_{0.0031}$ & $119_{3}$  \\
8  &   $350_{22}$  & $25828_{1801}$   & $0.66_{0.0153}$ & $0.070_{0.0020}$ & $133_{2}$  \\
10 &   $585_{22}$  & $41686_{1488}$   & $0.54_{0.0091}$ & $0.058_{0.0010}$ & $139_{1}$  \\
12 &   $896_{22}$  & $57420_{1012}$   & $0.46_{0.0048}$ & $0.052_{0.0006}$ & $140_{1}$  \\
13 &  $1085_{20}$  & $65287_{955}$    & $0.42_{0.0045}$ & $0.050_{0.0006}$ & $139_{1}$  \\

\hline \hline
\multicolumn{6}{|c|}{$N = 16$} \\
\hline
2  &     $33_{15}$ & $516_{358}$      & $0.96_{0.0245}$ & $0.326_{0.0579}$ & $56_{14}$   \\
4  &    $178_{33}$ & $9129_{2930}$    & $0.92_{0.0171}$ & $0.137_{0.0111}$ & $126_{8}$   \\
6  &    $460_{29}$ & $41791_{4690}$   & $0.79_{0.0154}$ & $0.084_{0.0028}$ & $170_{3}$   \\
8  &    $890_{33}$ & $93384_{4394}$   & $0.65_{0.0102}$ & $0.062_{0.0011}$ & $194_{2}$   \\
10 &  $1,470_{34}$ & $162139_{4592}$  & $0.53_{0.0070}$ & $0.050_{0.0006}$ & $206_{1}$   \\
12 &  $2,254_{32}$ & $227912_{2670}$  & $0.44_{0.0031}$ & $0.043_{0.0003}$ & $207_{1}$   \\
14 &  $3,264_{29}$ & $290732_{2056}$  & $0.38_{0.0022}$ & $0.040_{0.0003}$ & $203_{1}$   \\
15 &  $3,868_{33}$ & $321203_{2061}$  & $0.35_{0.0022}$ & $0.039_{0.0004}$ & $200_{1}$   \\

\hline \hline
\multicolumn{6}{|c|}{$N = 18$} \\
\hline
2  &     $50_{25}$ & $1579_{1854}$    & $0.95_{0.0291}$ & $0.307_{0.0630}$ & $73_{15}$   \\
4  &    $330_{72}$ & $26266_{7056}$   & $0.92_{0.0137}$ & $0.127_{0.0081}$ & $174_{9}$   \\
6  &    $994_{73}$ & $146441_{18685}$ & $0.78_{0.0155}$ & $0.076_{0.0044}$ & $237_{5}$   \\
8  &  $2,093_{70}$ & $354009_{18722}$ & $0.64_{0.0097}$ & $0.056_{0.0012}$ & $273_{2}$   \\
10 &  $3,619_{61}$ & $620521_{20318}$ & $0.52_{0.0071}$ & $0.044_{0.0007}$ & $292_{1}$   \\
12 &  $5,657_{59}$ & $899742_{14011}$ & $0.43_{0.0037}$ & $0.038_{0.0003}$ & $297_{1}$   \\
14 &  $8,352_{60}$ & $1163640_{11935}$& $0.36_{0.0023}$ & $0.034_{0.0002}$ & $293_{1}$   \\
16 & $11,797_{63}$ & $1406870_{6622}$ & $0.32_{0.0012}$ & $0.032_{0.0001}$ & $283_{1}$   \\
17 & $13,795_{77}$ & $1524730_{4818}$ & $0.30_{0.0009}$ & $0.032_{0.0001}$ & $277_{1}$   \\
\hline
\end{tabular}
\end{center}
\end{table}

\renewcommand{\baselinestretch}{2}

\subsection{Detailed Study of Network Features}

In this section we study in more depth some network features which
can be related to stochastic local search difficulty on the
underlying fitness landscapes. Table \ref{tab:statistics} reports
the average (over 30 independent instances for each $N$ and $K$) of
the network properties described. $\bar n_v$ and $\bar n_e$ are,
respectively, the mean number of vertices and the mean number of
edges of the graph for a given $K$ rounded to the next integer.
$\bar C^{w}$ is the mean weighted clustering coefficient. $\bar Y$
is the mean disparity, and $\bar d$ is the mean path length.

\subsubsection{Clustering Coefficients}

The fourth column of table \ref{tab:statistics} lists the average
values of the weighted clustering coefficients for all $N$ and $K$.
It is apparent that the clustering coefficients decrease regularly
with increasing $K$ for all $N$. For the standard unweighed
clustering, this would mean that the larger $K$ is, the less likely
that two maxima which are connected to a third one are themselves
connected. Taking weights, i.e.~transition probabilities into
account this means that either there are less transitions between
neighboring basins for high $K$, and/or the transitions are less
likely to occur. This confirms from a network point of view the
common knowledge that search difficulty increases with $K$.

\subsubsection{Shortest Path to the Global Optimum}

\begin{figure} [!ht]
\begin{center}
    \mbox{\includegraphics[width=8cm]{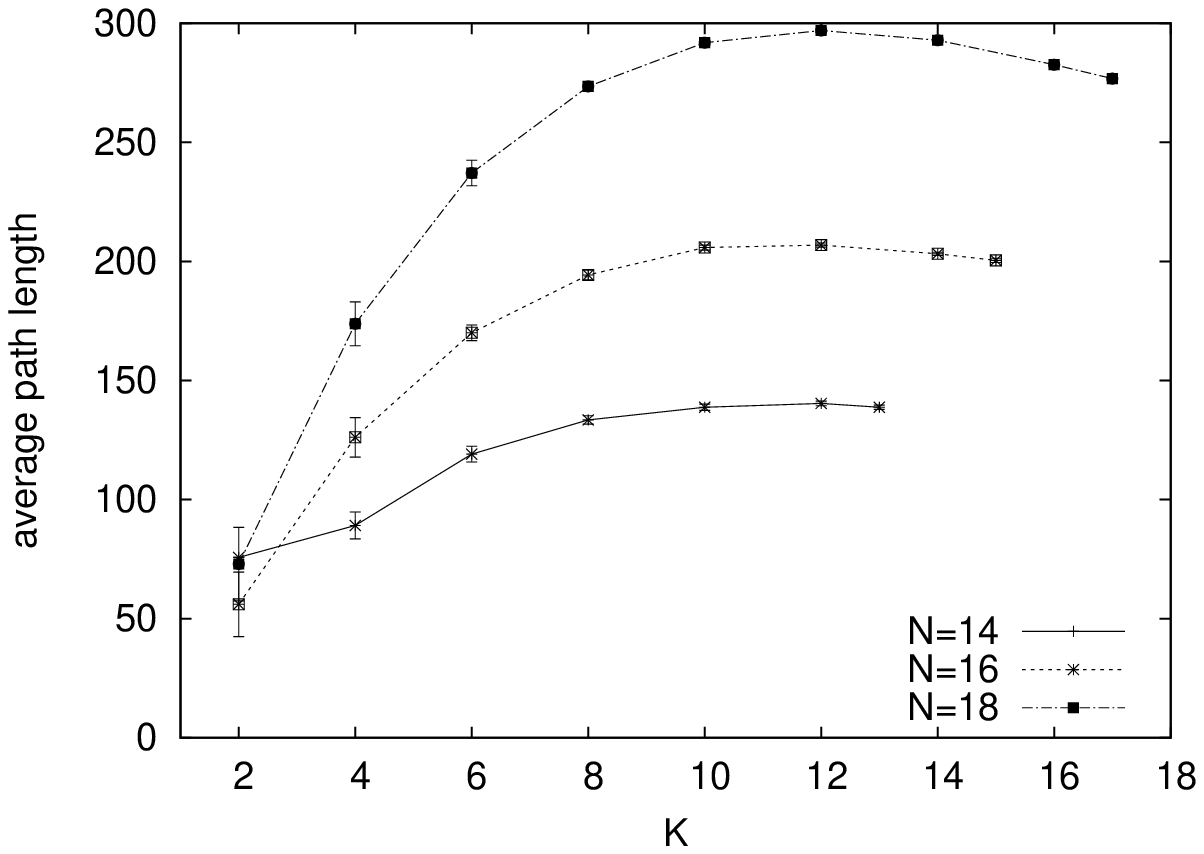} } \protect \\
    \mbox{\includegraphics[width=8cm]{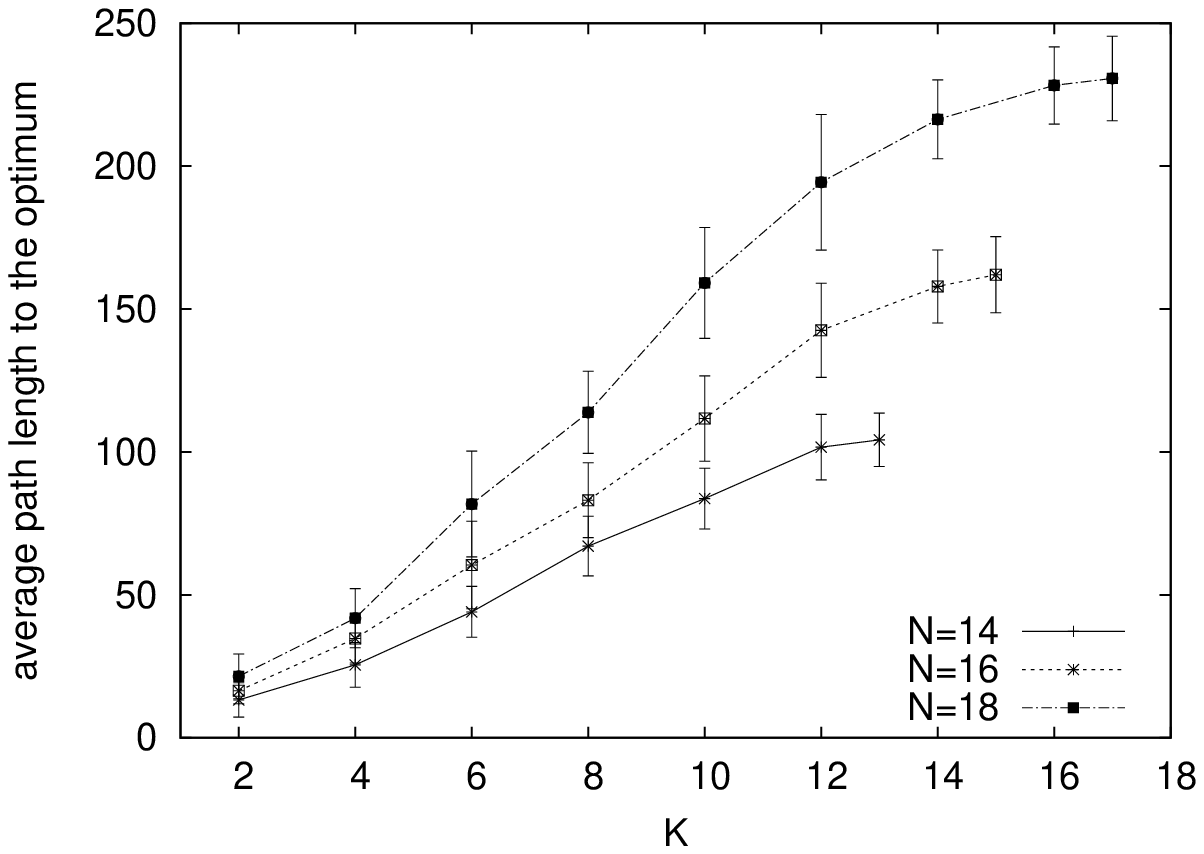} } \protect \\
    \vspace{-0.3cm}
\caption{Average distance (shortest path) between nodes (top), and
average path length to the optimum from all the other basins
(bottom). \label{fig:distances}}
\end{center}
\end{figure}

The average shortest path lengths $\bar d$ are listed in the sixth
column of table~\ref{tab:statistics}. Fig.~\ref{fig:distances} (top)
is a graphical illustration of the average shortest path length
between optima for all the studied $NK$ landscapes. Notice that the
shortest path increases with $N$, this is to be expected since the
number of optima increases exponentially with $N$. More
interestingly, for a given $N$ the shortest path increases with $K$,
up to $K = 10$, and then it stagnates and even decreases slightly
for the $N = 18$. This is consistent with the well known fact that
the search difficulty in $NK$ landscapes increases with $K$.
However, some paths are more relevant from the point of view of a
stochastic local search algorithm following a trajectory over the
maxima network. In order to better illustrate the relationship of
this network property with the search difficulty by heuristic local
search algorithms, Fig.~\ref{fig:distances} (bottom) shows the
shortest path length to the global optimum from all the other optima
in the landscape. The trend is clear, the path lengths to the
optimum increase steadily with increasing $K$.

\begin{figure}[!ht]
\begin{center}
\begin{tabular}{c}
{\tiny $N=16$}\\
\includegraphics[width=8cm]{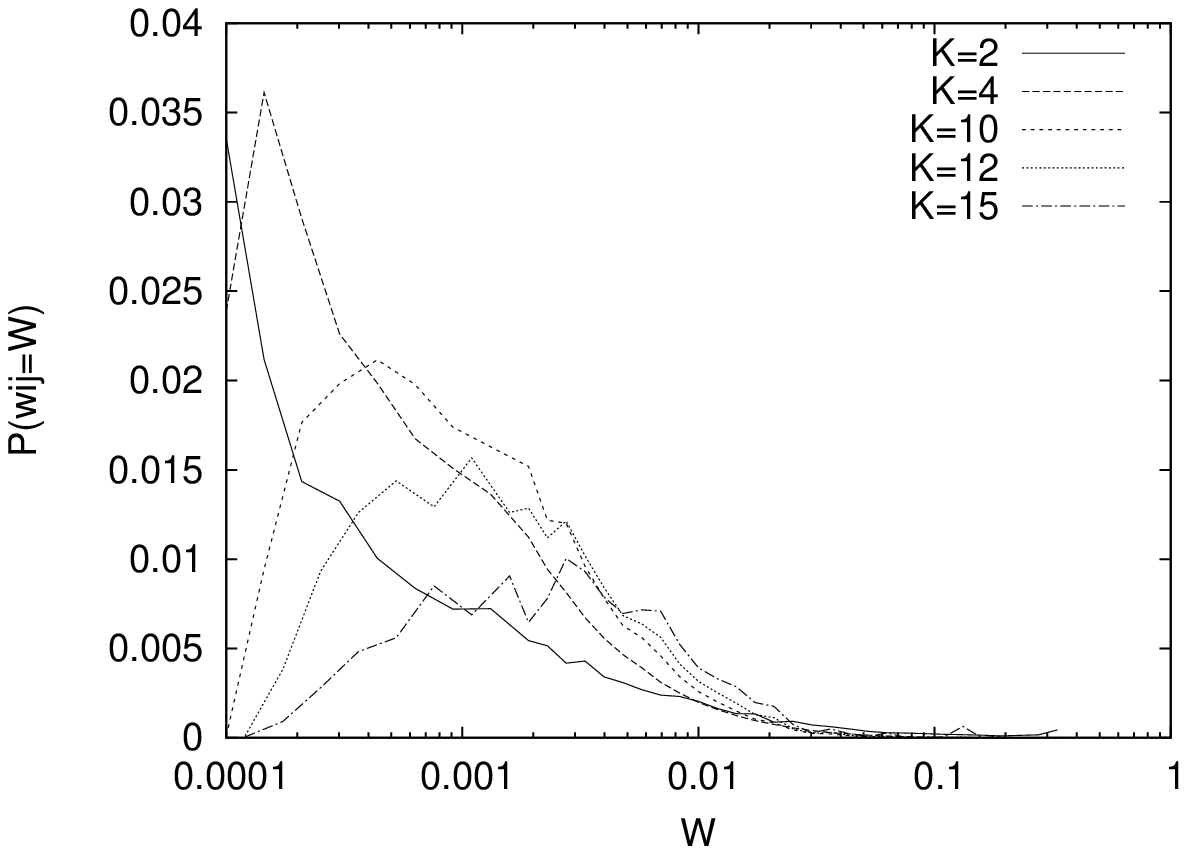}\\
{\tiny $N=18$}\\
\includegraphics[width=8cm]{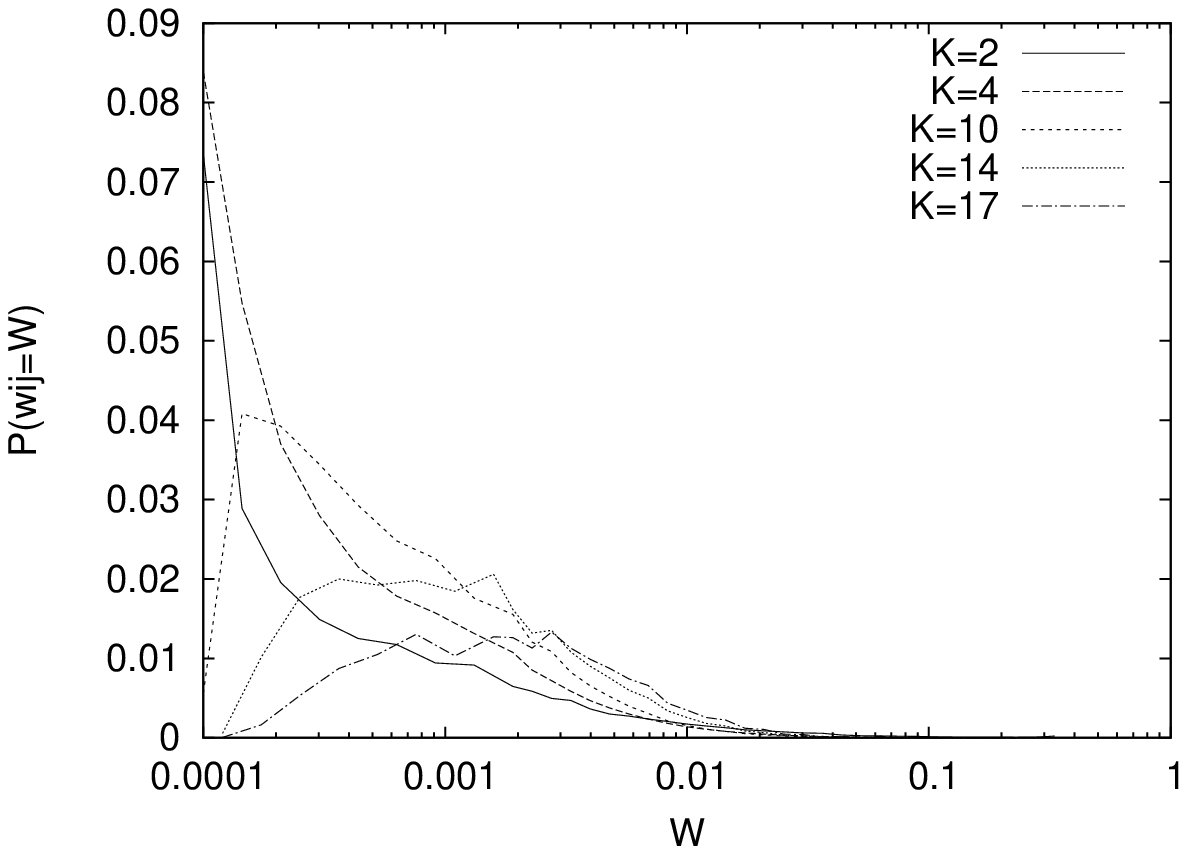} \\
\end{tabular}
\vspace{-0.3cm} \caption{Probability distribution of the network
weights $w_{ij}$ for outgoing edges with $j\not=i$ in log-scale on x-axis. Averages of
30 instances for each $N$ and $K$ are
reported.\label{fig-Distri-Wij}}
\end{center}
\end{figure}

\subsubsection{Outgoing Weight Distribution}

Here we report on the outgoing weight distributions $p_{out}(w)$ of
the maxima network edges. Fig.~\ref{fig-Distri-Wij} shows the
empirical probability distribution function for the cases $N=16$ and
$N=18$ (logarithmic scale has been used on the x-axis). The case
$N=14$ is similar but is not reported here because it is much more
noisy for $K=2$ and $4$ due to the small size of the graphs in these
cases (see table~\ref{tab:statistics}). One can see that the
weights, i.e. the transition probabilities to neighboring basins are
small. The distributions are far from uniform and, for both $N=16$
and $N=18$, the low $K$ have longer tails. For high $K$ the decay is
faster. This seems to indicate that, on average, the transition
probabilities are higher for low $K$.

We have already remarked that the approach taken
in~\cite{scala-et-al-01,doye02,doye05} is different in that edges
between two optima either exist or not; in other words the notion of
transition probability is absent.  However, it is worth recalling
that Doye et al.~\cite{doye02,doye05} found that their inherent
networks were of the scale-free type with the global minimum being
often the most connected node. The landscape is thus essentially a
``funnel'' meaning that the system can relax to the global minimum
from almost any other local minimum. In our language, we would say
that this kind of landscape is an `easy'' one to search. In
contrast, the inherent networks found by Scala et
al.~\cite{scala-et-al-01} for lattice polymer chains are of the
small-world type but they show a fast-decaying degree distribution
function.

\subsubsection{Disparity}

\begin{figure} [!ht]
\begin{center}
\begin{tabular}{c}
     {\tiny $N=16$} \\
    \mbox{\includegraphics[width=8cm]{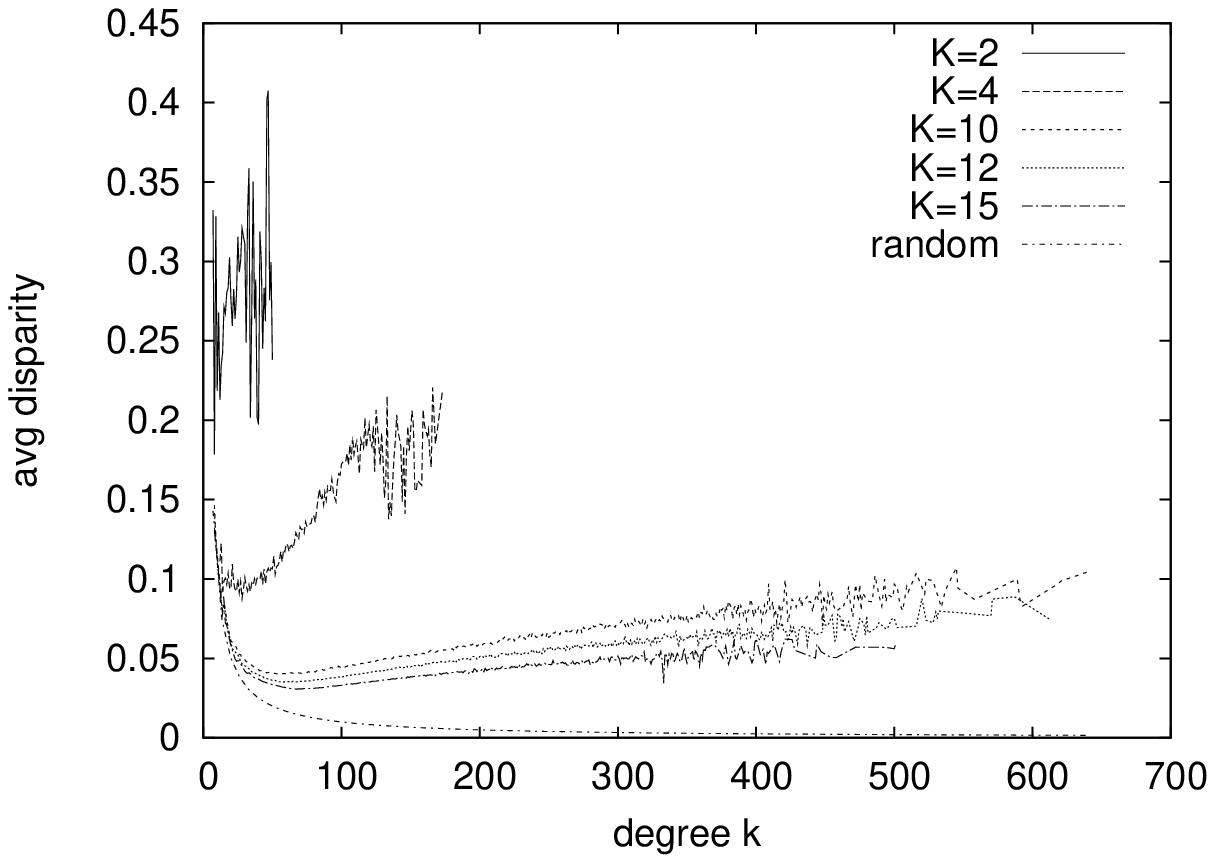}}\\
    {\tiny $N=18$} \\
    \mbox{\includegraphics[width=8cm]{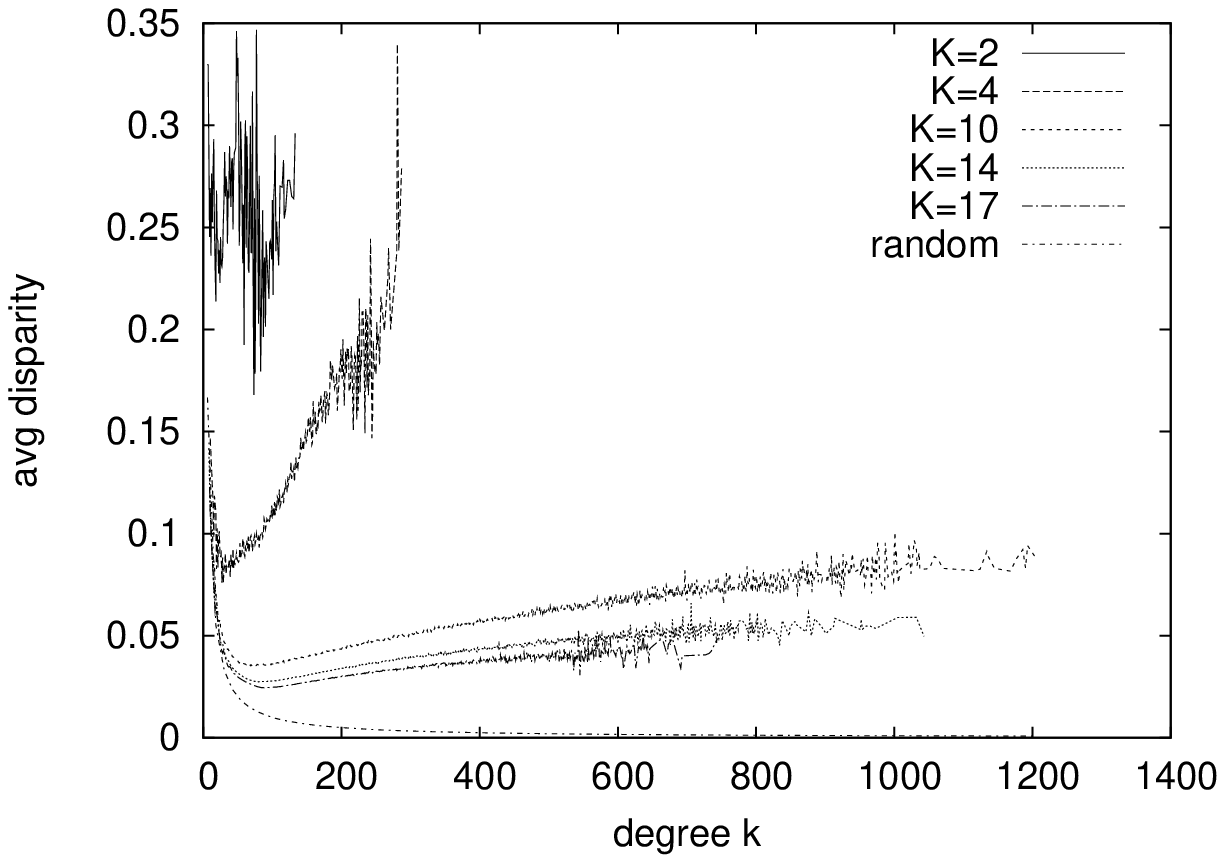}}\\
\end{tabular}
\vspace{-0.3cm} \caption{Average disparity,  $Y_{2}$, of nodes with
a given degree $k$. Average of $30$ independent instances for each
$N$ and $K$ are reported. The curve $1/k$ is also reported to
compare to random case.\label{fig-deg-disparity}}
\end{center}
\end{figure}

Fig.~\ref{fig-deg-disparity} depicts the disparity coefficient as
defined in the previous section for $N=16, 18$. An interesting
observation is that the disparity (i.e.~inhomogeneity) in the
weights of a node's out-coming links tends to decrease steadily with
increasing $K$. This reflects that for high $K$ the transitions to
other basins tend to become equally likely, which is another
indication that the landscape, and thus its representative maxima
network, becomes more random and difficult to search.

When $K$ increases, the number of edges increases and the number of
edges with a weight over a certain threshold increases too (see
fig.~\ref{fig-Distri-Wij}). Therefore, for small $K$, each node is
connected with a small number of nodes each with a relative high
weight. On the other hand,  for large $K$, the weights become more
homogeneous in the neighbourhood, that is, for each node, all the
neighboring basins are at similar distance.

If we make the hypothesis that edges with higher weights are likely to be
connected to nodes with larger basins (an intuition that we need to
confirm in future work) then, as the larger basins tend to have
higher fitness (see Fig.~\ref{fig:cor_fit-size}), the path to higher
fitness values would be easier to find for lower $K$ than for larger
$K$.

\subsubsection{Boundary of basins}

\begin{figure} [!ht]
\begin{center}
    \mbox{\includegraphics[width=8cm]{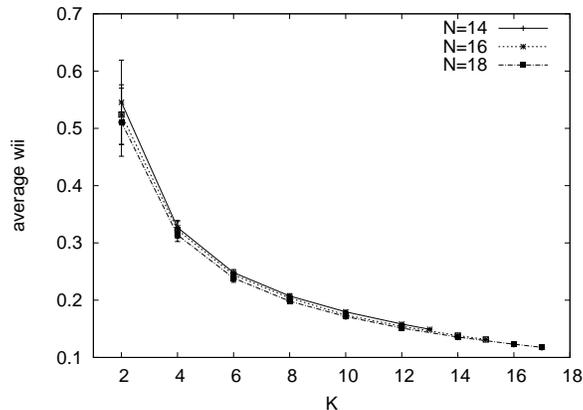} } \protect \\
\vspace{-0.3cm} \caption{Average weight $w_{ii}$ according to the
parameters $N$ and $K$. \label{fig:wii}}
\end{center}
\end{figure}

Fig. \ref{fig:wii} shows the averages, over all the nodes in the
network, of the weights $w_{ii}$ (i.e. the probabilities of remaining
in the same basin after a bit-flip mutation). Notice that the
weights $w_{ii}$ are much higher when  compared to those $w_{ij}$
with $j \not= i$ (see  Fig.~\ref{fig-Distri-Wij}). In particular,
for $K=2$, $50\%$ of the random bit-flip mutations will produce a
solution within the same basin of attraction. These average
probabilities of remaining within the same basin, are  above $12 \%$
for the higher values of $K$. Notice that the averages are nearly
the same regardless the value of $N$, but decrease with the
epistatic parameter $K$.

The exploration of new basins with the random bit-flip mutation
seems to be, therefore, easier for large $K$ than for low $K$. But,
as the number of basins increases, and the fitness correlation
between neighboring solutions decreases with increasing $K$, it
becomes harder to find the global maxima for large $K$. This result
suggests that the dynamic of stochastic local search algorithms on
$NK$ landscapes with large $K$ is different from that with lower
values of $K$, with the former engaging in more random exploration
of basins.

Table~\ref{tab:inner} gives the average number of configurations in
the interior of basins (this statistic is computed on $30$
independent landscapes). Notice that the size of the basins'
interior is below $1\%$ (except for $N=14$, $K=2$). Surprisingly,
the size of the basins' boundaries is nearly the same as the size of
the basins themselves. Therefore, the probability of having a
neighboring solution in the same basin is high, but nearly all the
solutions have a neighbor solution in another basin. Thus, the
interior basins seem to be \textit{``hollow"}, a picture which is
far from the smooth standard  representation of landscapes in 2D
with real variables where the basins of attraction are visualized as
real mountains.

\begin{table}
\vspace{-0.3cm} \caption{Average (on $30$ independent landscapes for
each $N$ and $K$) of the mean relative sizes of the basins
interiors.\label{tab:inner}}
\begin{center}
\begin{tabular}{|l|l|l|l|}
\hline
K & $N=14$ & $N=16$ & $N=18$ \\
\hline
2   & $0.0167_{0.02478}$ & $0.0050_{0.00798}$ & $0.0028_{0.00435}$ \\
4   & $0.0025_{0.00065}$ & $0.0012_{0.00025}$ & $0.0006_{0.00010}$ \\
6   & $0.0029_{0.00037}$ & $0.0014_{0.00015}$ & $0.0007_{0.00009}$ \\
8   & $0.0043_{0.00045}$ & $0.0022_{0.00012}$ & $0.0011_{0.00006}$ \\
10  & $0.0055_{0.00041}$ & $0.0031_{0.00015}$ & $0.0018_{0.00006}$ \\
12  & $0.0061_{0.00041}$ & $0.0040_{0.00014}$ & $0.0025_{0.00007}$ \\
13  & $0.0059_{0.00029}$ &                   &                   \\
14  &                   & $0.0045_{0.00012}$ & $0.0031_{0.00004}$ \\
15  &                   & $0.0044_{0.00014}$ &                   \\
16  &                   &                   & $0.0035_{0.00005}$ \\
17  &                   &                   & $0.0034_{0.00006}$ \\
\hline
\end{tabular}
\end{center}
\end{table}
\vspace{-0.3cm}

\subsubsection{Incoming Weights Distribution}
\label{in-distr}
\begin{figure} [!ht]
\begin{center}
\includegraphics[width=8cm] {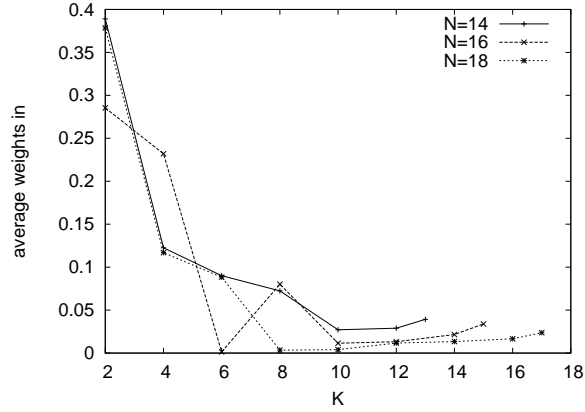} \protect \\
\caption{Average value of the weights of incoming transitions into maxima nodes for
$N=14,16,18$ and for whole $K$ interval.\label{av-into}}
\end{center}
\end{figure}

\begin{figure*} [!ht]
\begin{center}
\begin{tabular}{cccc}
    \mbox{\includegraphics[width=6.3cm]{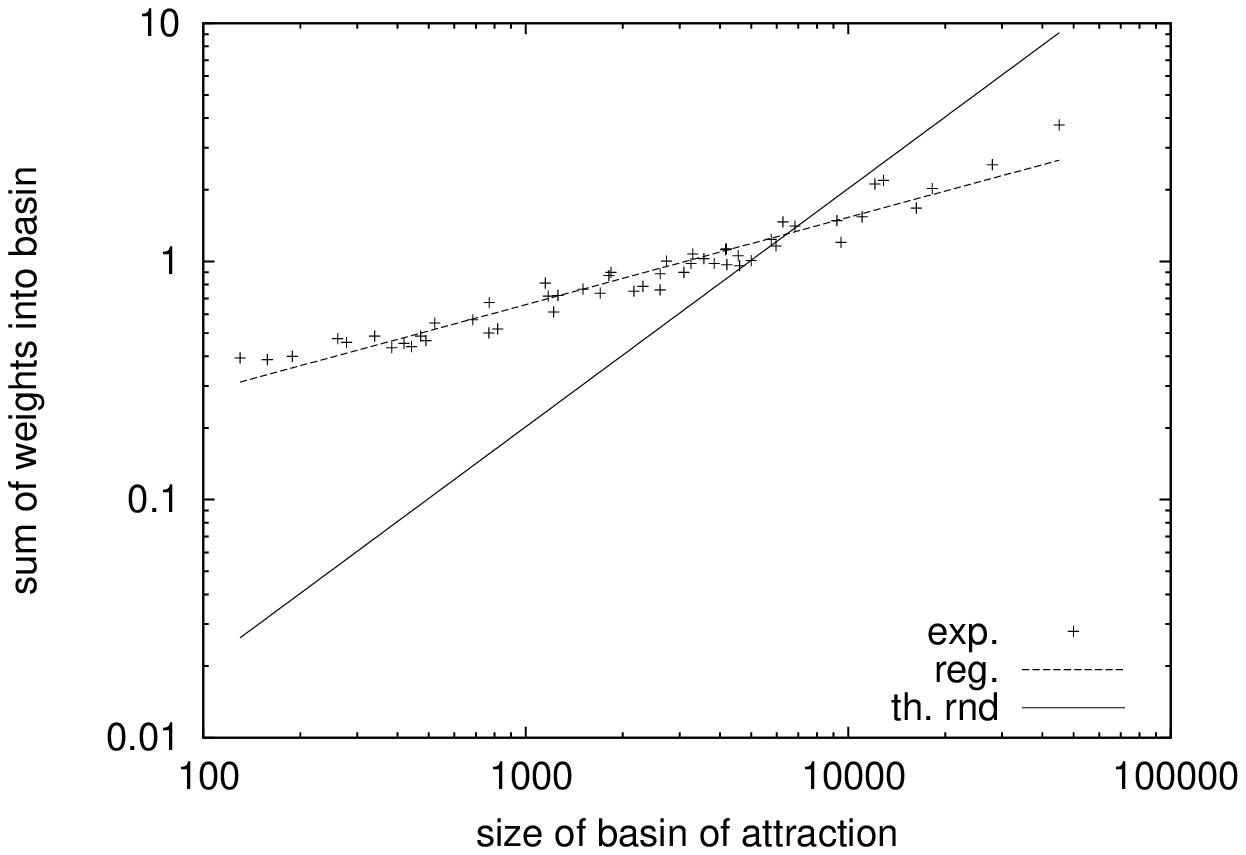} } \protect &
    \mbox{\includegraphics[width=6.3cm]{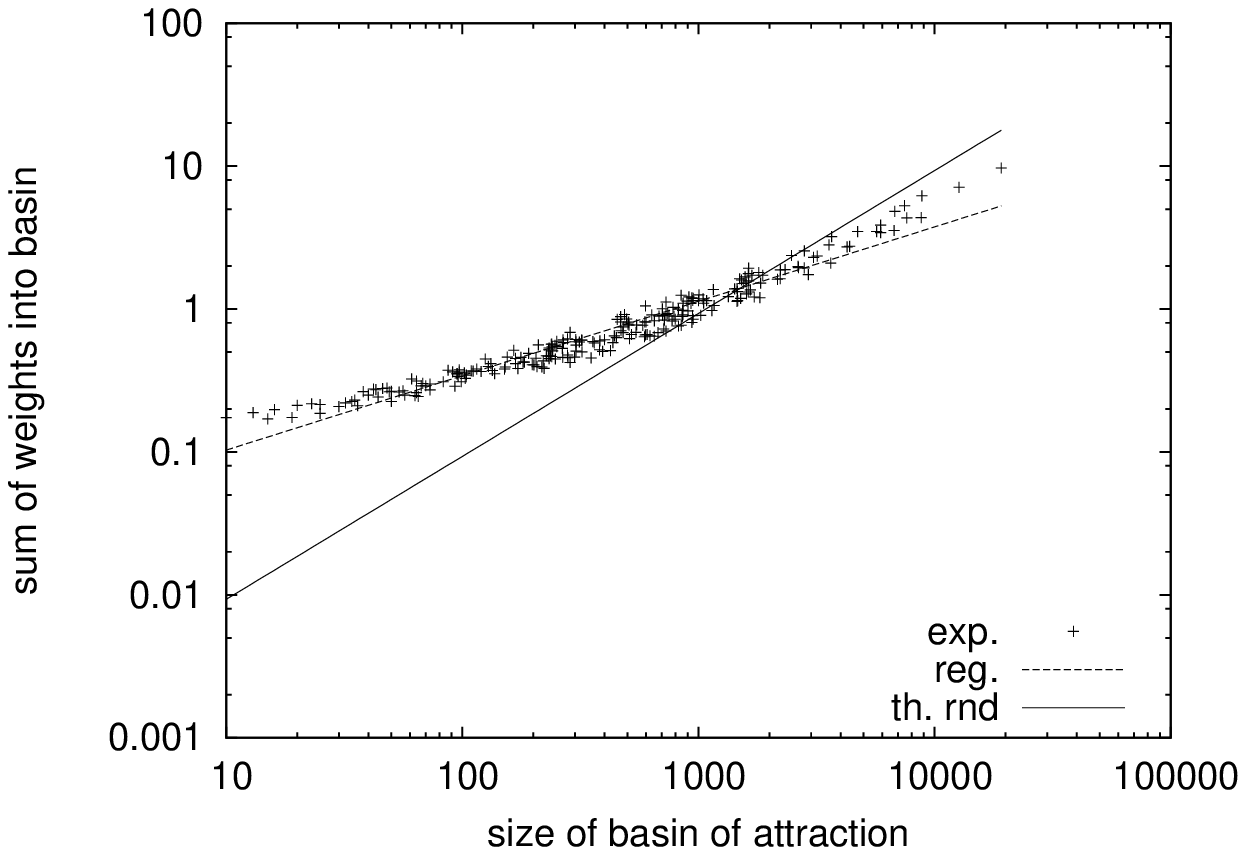} } \protect \\
    K=2 & K=4\\
    \mbox{\includegraphics[width=6.3cm]{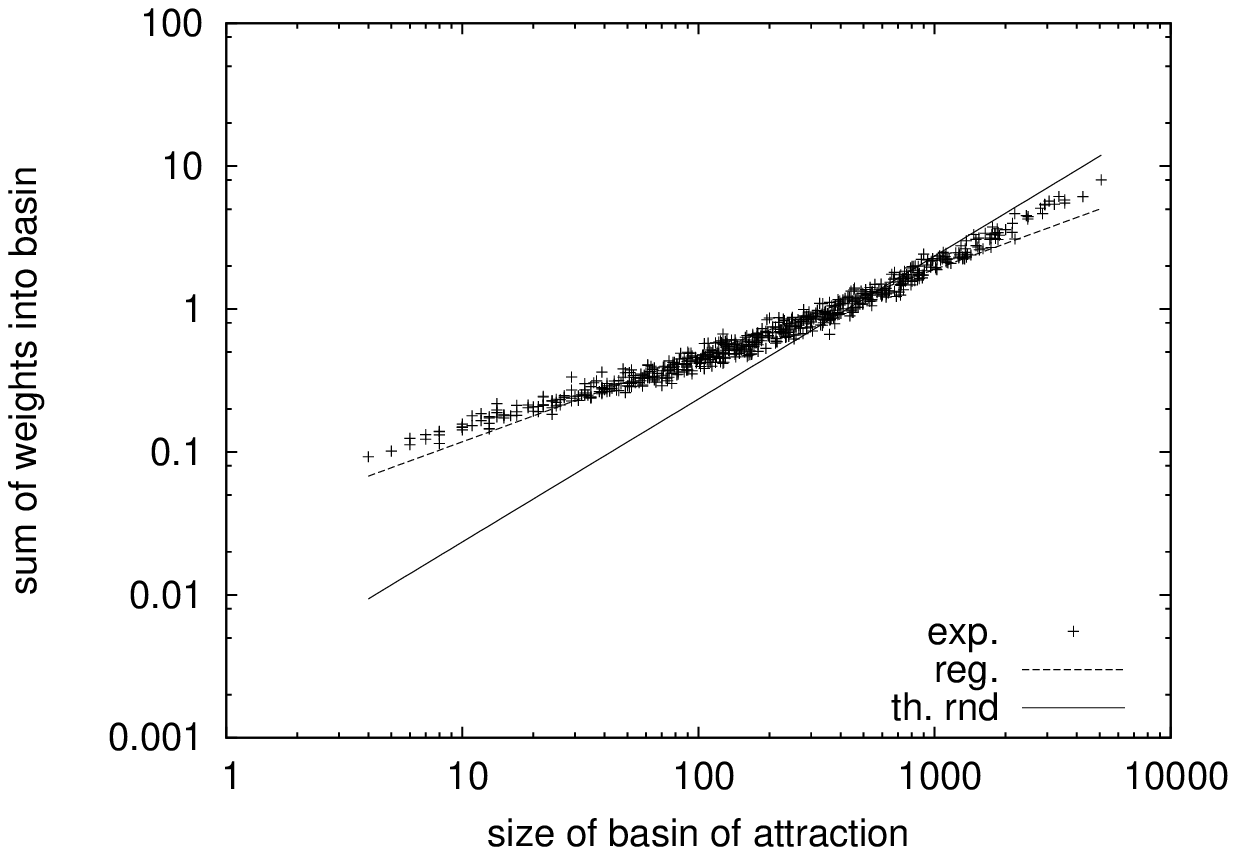} } \protect &
    \mbox{\includegraphics[width=6.3cm]{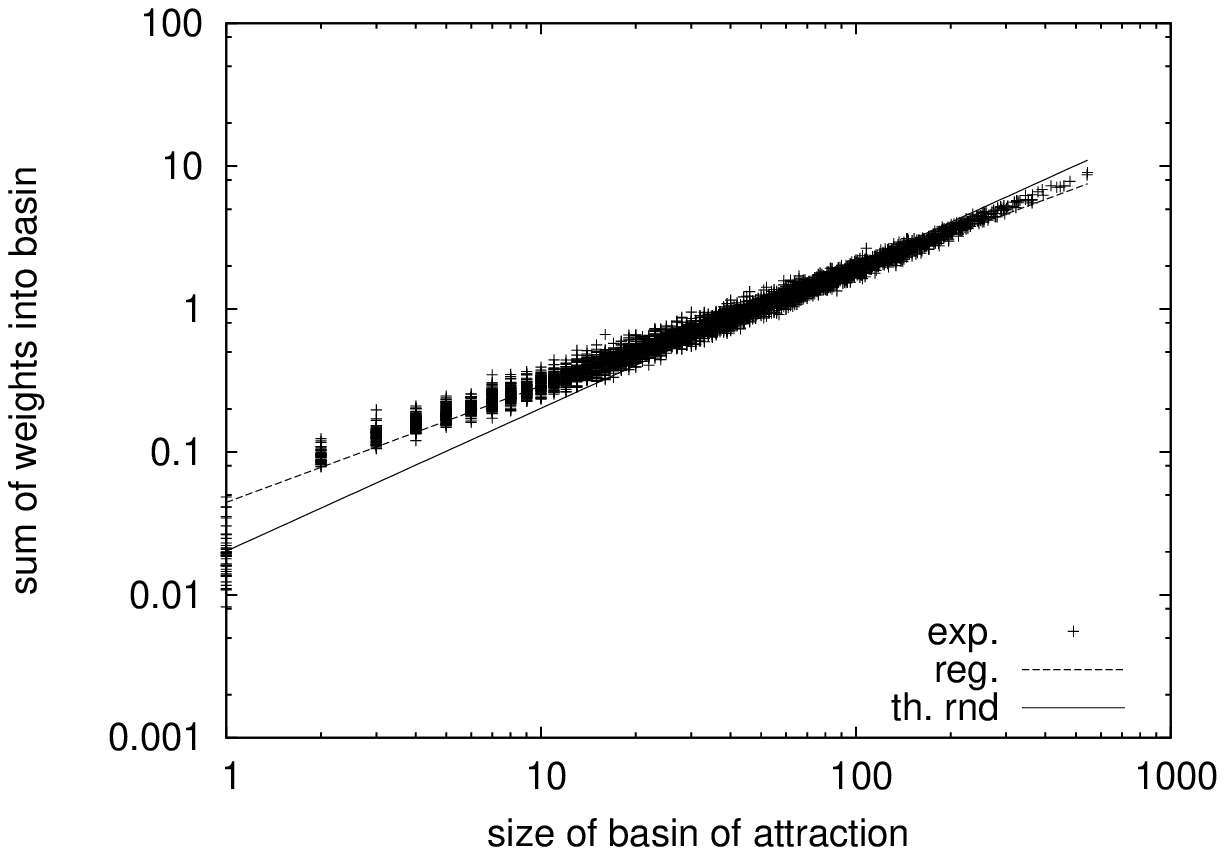} } \protect \\
    K=6 & K=12
    \vspace{-0.3cm}
 \end{tabular}
\caption{Scatter-plots of the sum of weights into basins vs. the
basin size for $N=18$ and various $K$ values. Log-log scale. The
regression line of each scatter-plot is shown. The theoretical
curves, under the hypothesis that the weights are proportional, to
the size of basin are also given. \label{bscorr}}
\end{center}
\end{figure*}

It is also of interest to study the  distribution of the weights of edges
impinging into a given node $p_{in}(w)$. However, a plot of this quantity is not very
informative. We prefer to show in Fig.~\ref{av-into} the average values over $30$ independent
lansdcapes for each value of $N$ and for the whole interval of $K$.
The general trend for all values of $N$ is that the average weight
of the incoming transitions into a node quickly decreases with
increasing $K$. This means that it is more difficult to make a
transition to a local maximum when $K$ is large. This agrees with
the fact that the relative basins' size is a rapidly decreasing
function of $K$ (see Fig.~\ref{nbasins}). In fact, there is a strong
positive correlation between the basins' size and the weights of the
transitions into the corresponding maximum, i.e. as the basin
becomes larger, the number of transitions into it increases too.
This is shown on the scatter-plots (Fig.~\ref{bscorr}). The
correlation follows approximately a power-law, the regression lines
are also visualized. The correlation coefficient for all plots is
high (above $0.97$). If we hypothesize that the incoming weights are
proportional to the size of the basin, i.e. that edges between nodes
are randomly distributed over the search space, the sum of the
incoming weights for a basin $b_i$ should be $\frac{\sharp b_i}{2^N}
\sharp S^{*}$. This theoretical line is visualized on the
scatter-plots (Fig.~\ref{bscorr}). Notice that the difference
between the theoretical and regression lines, is higher for low
values of $K$. For large $K$, the weights are given almost only by
the size of basins. This is not the case for small $K$ values, where
the fitness correlation between neighboring solutions is high
\cite{weinberger91}. This explains why the hypothesis does not hold
in this case. So, the incoming weights could be deduced from the
size of basins and the fitness of solutions belonging to the basin.


\section{Conclusions}
\label{conclusions}

We have proposed a new characterization of combinatorial fitness
landscapes using the well-known family of $NK$ landscapes as an
example. We have used an extension of the concept of inherent
networks proposed for energy surfaces~\cite{doye02} in order to
abstract and simplify the landscape description. In our case the
inherent network is the graph where the nodes are all the local
maxima and the edges accounts for transition probabilities (using
the 1-flip operator) between the local maxima basins of attraction.
This mapping leads to oriented weighted graphs, instead of the more
commonly used unordered and unweighed ones~\cite{doye02}. We believe
that the present representation is closer to the view offered by a
Monte Carlo search heuristic such as simulated annealing which
produces a trajectory on the configuration space based on transition
probabilities according to a Boltzmann equilibrium
distribution\cite{KGV-SA}. We have exhaustively obtained these
graphs for $N=\{14, 16, 18\}$, and for all even values of $K$, plus
$K=N-1$, and conducted a network analysis on them.  The network
representation of the $NK$ fitness landscapes has proved useful in
characterizing the topological features of the landscapes and gives
important information on the structure of their basins of
attraction. In fact, our guiding motivation has been to relate the
statistical properties of these networks, to the search difficulty
of the underlying combinatorial landscapes when using stochastic
local search algorithms (based on the bit-flip operator) to optimize
them.
 We have found clear indications of such
relationships, in particular:

\begin{description}
\item [The clustering coefficients:] suggest that, for high values of $K$,
the transition between a given pair of neighboring basins is less
likely to occur.
\item [The shortest paths to the global optimum:] become longer with
increasing $N$, and for a given $N$, they clearly increase with
higher $K$.
\item [The outgoing weight distribution:] indicate that, on average, the
transition probabilities from a given node to neighbor nodes are higher for low $K$.
\item [The incoming weight distribution:] indicate that, on average, the
transition probabilities from the neighborhood of a node become lower  with
increasing $K$.
\item [The disparity coefficients:] reflect that for high $K$ the transitions to
other basins tend to become equally likely, which is an indication
of the randomness of the landscape.
\end{description}

The previous results clearly confirm and  justify from a novel
network point of view the empirically known fact that $NK$ landscapes become
harder to search as they become more and more random with increasing $K$.

The construction of the maxima networks requires the determination
of the basins of attraction of the corresponding landscapes. We have
thus also described the nature of the basins, and found  that the
size of the basin corresponding to the global maximum becomes
smaller with increasing $K$. The distribution of the basin sizes is
approximately exponential for all $N$ and $K$, but the basin sizes
are larger for low $K$, another indirect indication of the
increasing randomness and difficulty of the landscapes when $K$
becomes large. Furthermore, there is a strong positive correlation
between the basin size  and the degree of the corresponding maximum,
which confirms that the synthetic view provided by the maxima graph
is a useful. Finally, we found that the size of the basins
boundaries is roughly the same as the size of basins themselves.
Therefore, nearly all the configurations in a given basin have a
neighbor solution in another basin. This observation suggests a
different landscape picture than the smooth standard representation
of 2D landscapes where the basins of attraction are visualized as
real mountains. Some of these results on basins in $NK$ landscapes
were previously unknown~\cite{kauffman93}.

This study represents our first attempt towards a topological and
statistical characterization of easy and hard combinatorial
landscapes, from the point of view of complex networks analysis.
Much remains to be done. First of all, the results found should be
confirmed for larger instances of $NK$ landscapes. This will require
good sampling techniques, or theoretical studies since exhaustive
sampling becomes quickly impractical. Other landscape types should
also be examined, such as those containing neutrality, which are
very common in real-world applications, and especially the
landscapes generated by important hard combinatorial problems such
as the traveling salesman problem and other resource allocation
problems. Work is in progress for neutral versions of $NK$
landscapes and for knapsack problems. Finally, the landscape
statistical characterization is only a step towards implementing
good methods for searching it. We thus hope that our results will
help in designing or estimating efficient search techniques and
operators.

\renewcommand{\baselinestretch}{1}       
\small
\bibliographystyle{unsrt}


\end{document}